\documentclass[12pt,aps,floats]{revtex4}
\usepackage{graphicx}
\usepackage{epsfig}

\def\laq{\ \raise 0.4ex\hbox{$<$}\kern -0.8em\lower 0.62 ex\hbox{$\sim$}\ }
\def\gaq{\ \raise 0.4ex\hbox{$>$}\kern -0.7em\lower 0.62 ex\hbox{$\sim$}\ }
\newcommand{\cN}{{\cal N}}

\begin{document}
\title{M-Theory Moduli Space and Cosmology}

\author{R. Brustein\protect\( ^{(1)}\protect \),
S. P. de Alwis\protect\( ^{(2)}\protect \), E. G. Novak
\protect\(^{(1)}\protect \)}

\address{(1) Department of Physics, Ben-Gurion University,
Beer-Sheva 84105, Israel
\\
 (2) Department of Physics, Box 390, University of Colorado,
 Boulder, CO 80309.\\
 \texttt{e-mail:  ramyb@bgumail.bgu.ac.il },
\texttt{dealwis@pizero.colorado.edu},
\texttt{enovak@bgumail.bgu.ac.il}}

\begin{abstract}
We conduct a systematic search for a viable string/M-theory
cosmology, focusing on cosmologies that include an era of
slow-roll inflation, after which the moduli are stabilized and
the Universe is in a state with an acceptably small
cosmological constant. We observe that the duality relations
between different cosmological backgrounds of string/M-theory
moduli space are greatly simplified, and that this
simplification leads to a truncated moduli space within which
possible cosmological solutions lie.  We review some known
challenges to four dimensional models in the ``outer",
perturbative, region of moduli space, and use duality relations
to extend them to models of all of the (compactified) 
perturbative string theories and 11D supergravity, including
brane world models. We conclude that cosmologies restricted to
the outer region are not viable, and that the most likely
region of moduli space in which to find realistic cosmology is
the ``central", non-perturbative region, with coupling and
compact volume both of order unity, in string units. 
\end{abstract} 
\maketitle

\section{introduction}

Cosmology offers a unique opportunity for testing string theory
models. However, a prerequisite for taking advantage of this
opportunity is a consistent framework that is not in obvious
contradiction with basic known features about cosmology and
particle phenomenology. This realization has revived interest in
classifying time-dependent/cosmological string backgrounds and
determining which class of solutions may be used to construct a
viable string cosmology.

The growing flow of increasingly accurate data from cosmological
observations, in addition to upcoming experiments with the power
to discriminate between cosmological models offer the hope of
guiding theory and confronting its predictions with accurate
measurements. In particular, an epoch of primordial slow-roll
inflationary evolution seems to be the simplest possible
explanation for the existing data on cosmic microwave background (CMB) 
anisotropies and large
scale structure observations (see for example,
\cite{Bond:2002cg}); thus it is desirable to incorporate it into
models of string cosmology (see, for example,
\cite{Linde:2001qe}).

With this widespread interest in the predictions of string
cosmology, one seems to be presented with a bewildering array of
models to choose from, that seemingly could be shaped to fit any
desirable results. However, there are well-known generic
obstacles to obtaining slow-roll inflation, and to stabilizing
moduli after inflation in a state with an acceptably small
cosmological constant \cite{BS}. These obstacles were originally discovered
in the context of perturbative heterotic string theory,
and could have been perceived as specific to that particular
context. Since current cosmological models of other string
theories, and braneworld scenarios based on them
(see, for example,  \cite{Quevedo:2002xw,Ovrut:2002hi}) have gained
popularity, it might have been assumed that in other string
theories such problems are more easily overcome. However, the
lesson of string duality is that there is a nice cohesion to the
physics of string theory \cite{Polchinski}.
Dualities relate all the perturbative
string theories and 11D supergravity (SUGRA). The strongly
coupled limit of one particular theory, rather than being
unknown, is something familiar, especially in the low energy
limit described by SUGRA.

In this paper we will use the duality relations among the
different corners of moduli space of supersymmetric vacua in
string/M-theory to relate different models of string cosmology.
We will be able to examine which regions of the moduli space lead
to unsatisfactory cosmologies suffering from the aforementioned
cosmological problems, and which regions may lead to promising
cosmology.   Our hope is that with this tool in hand, it will be
easier to isolate and probe the sorts of background manifolds
that will have the most promising behavior. Some preliminary
progress in this direction has already been achieved
\cite{center}, and we use it to highlight the essential features
of a viable string cosmology.

We should mention that the most detailed approach to constructing a
consistent string cosmology is to give up on having a slow-roll
phase of inflation and rather have ``fast-roll" inflation in
the pre-history of the Universe \cite{PBB}. Fast-roll inflation
is more sensitive to the details of the cosmological evolution, as
opposed to the ``no-hair" nature of slow-roll inflation. If the
most recent period of inflation was of slow-roll type, then the
preceding ``pre-history" is hidden from us, so in this case it is
sufficient to study cosmology from the last phase of slow-roll
inflation and on, which is the attitude that we have adopted in
this paper.

We begin in the next section by describing the action of duality
in a cosmological context, making the observation that within the
assumptions of string cosmology, the moduli space of solutions
and the duality actions upon it greatly simplify.  In section
III, we divide the moduli space into ``safe'' regions where we
trust perturbation theory, ``unsafe'' regions where we don't, and
a ``central'' region which is in some sense maximally unsafe.  We
then see how the different regions transform under duality, and
make some preliminary arguments about physics in the central
region.  We continue in this section by looking at specific
solutions to the cosmological equations of motion, and
demonstrate how the various duality transformations act on these
solutions. The explicit transformations can be found in the appendix. We
conclude this section with a discussion of the
physics of solutions in the outer region of moduli space.

In section IV, we extend our analysis to the case that there are
non-perturbative potentials for the moduli, for example from
brane instantons. Here we argue that the presence of potentials
strengthens our argument for the importance of the central
region.   In section V we translate our analysis into the
framework of various brane world cosmologies, and make some
comments on the physics of these models.  In section VI we
briefly discuss cosmology in the central region, summarizing the
argument made in \cite{center}, and we end with some conclusions
and future directions in section VII.

\section{string cosmology and duality}

We begin this section by explicitly stating  the assumptions  on
which we base our approach to string cosmology.  We shall
continue by looking at how string duality transformations work
within these assumptions, giving simple, explicit transformations
that we can apply to cosmological solutions.

We are interested in finding cosmological solutions of
String/M-theory that will describe homogeneous and isotropic 4
dimensional space-times. Our primary interest will be to isolate
any solutions that produce a slow-roll inflationary era ending
with stabilized moduli in a universe with an acceptably small
cosmological constant. Of course, these solutions will also have
the usual extra, compact dimensions, which are necessary for the
consistency of the theory. Although we are far from having
satisfactory control over the full, non-perturbative completion
of string theory, we know that its low energy effective field
theory should be described by SUGRA. In particular, the energy
scale of reasonable inflaton potentials responsible for slow-roll
inflation is generally far smaller than the Planck scale. SUGRA
should therefore describe the cosmological evolution very well
throughout an inflationary era and for all but the very highest
curvatures.

What differentiates our analysis from a pure SUGRA analysis,
however, is that we always have in mind a derivation from string
theory.  Thus, we only trust our solutions when we trust their
stringy completion.  In particular, we trust our solutions only
when they correspond to a string background with small string
coupling and low string-frame space-time curvature.

There are three key features exhibited by such stringy
cosmological solutions. The first is that string theory
phenomenology prefers the use of SUGRA for the low energy
effective theory, rather than simple gravity, and we are forced
to address the problem of moduli such as the dilaton.  Any
regions where we trust perturbative string theory will have these
moduli, and in particular we must deal with the well known
instability, first elucidated by Dine and Seiberg \cite{DS},
which drives perturbative strings to the free, noncompact limit,
and its cosmological counterpart \cite{BS}. Many studies of
string cosmology make the assumption that there is a
non-perturbative mechanism for stabilizing moduli whose effects on
cosmological physics are subsequently ignored.  In this work,
instead, we find that the mechanism of moduli stabilization can
be a natural source for realistic, slow-roll inflation. A second
feature is that since we have forced ourselves to accept some of
the difficulties of string theory, and have been careful that the
embedding is justified, when the solution drives the Universe
towards a state with high energy densities, we will assume that
in some cases string theory can make some sense of it. Finally,
we note that by embedding our solutions in the framework of
string theory, we can use string dualities to our advantage.  The
structure of the moduli space of string theory supplies our model
with a valuable framework, within which we can exercise greater
control over the necessary approximations.  In particular, if the
gravity solution goes to a strong-coupling or strong (compact)
curvature regime, we can use string duality to map our solution
to a \emph{different} solution whose physics is often well-controlled.

In this paper we focus on Einstein gravity coupled to scalar
moduli fields in 4 space-time dimensions. In general, these
fields are imbedded within an $N=1$ SUGRA theory, whose full form
is dictated by string theory. Most of the details of the SUGRA
theory are unimportant for cosmology, apart from a possible
scalar potential, whose effects we will consider in section IV.
Thus we use the simple lagrangian:
 \begin{equation}
S_4 = \frac{1}{2 \kappa_4^2} \int d^4x\sqrt{-g}[R - \frac{1}{2} g^{\mu \nu}
\partial_\mu \phi \partial_\nu \phi -\frac{1}{2}g^{\mu \nu}
\partial_\mu \sigma \partial_\nu \sigma -V(\phi, \sigma)] \quad .
 \label{eq:grav_action}
 \end{equation}
The fields $\phi$ and $\sigma$ are two representative moduli fields that
control the string coupling and, in our example, the size of the
compact space. This restriction on models comes about because we
are interested in spatially homogeneous and isotropic 4
dimensional spacetimes. The scalar moduli serve as the most
likely inflaton candidates in string cosmology \cite{BG}.

Recall that the action (\ref{eq:grav_action}) is a restriction of
the full low-energy effective SUGRA action.  As such, it inherits
dualities from the full string theory.  There is, of course a
menagerie of S, T, and U string dualities, as well as M-theory
strong coupling limits where the space-time grows extra
dimensions.  These different dualities have very distinct and
specific actions on string states that can be very complicated.
When restricted to dilaton gravity, however, the dualities
simplify considerably.  For example, we may consider T-duality,
which depends crucially on the string winding states that are
interchanged with Kaluza-Klein modes along the compact
directions.  Since our cosmological models have no need to pay
attention to the detailed dynamics of momentum or winding states,
the only effect of T-duality is an inversion of the size of the
compactification and a shift in the dilaton.  Likewise, the
evidence for S-duality rests largely on the behavior of the
spectrum of BPS states, which are irrelevant for cosmology. Thus,
general duality transformations can be applied in a simplified
way to extend the range of trustworthy solutions, a procedure we
shall now illustrate in more detail. Dualities were discussed in
the context of string cosmology in \cite{coprev}, with emphasis
on the symmetries of cosmological solutions of a single low
energy effective action. Here our focus is on the relations
between cosmological solutions of different string theories.

Our 4D lagrangian comes from a dimensional reduction of the 10D
action. Restricting ourselves to the gravity plus dilaton sector
of the theory, all of the low energy effective actions for the
different perturbative string theories reduce to the same simple
form:
 \begin{equation}
S_{10} = \frac{1}{2 \kappa_{10}^2} \int d^{10}x
\sqrt{-G}e^{-2\Phi}\left(R_{10} + 4 G^{MN}\partial_M \Phi
\partial_N
\Phi \right) \quad ,
 \end{equation}
where $G_{MN}$ is the ten dimensional string-frame metric of the
corresponding string theory, and $\Phi$ is the ten dimensional
dilaton of that theory. For simplicity, we shall initially assume
that 6 directions are toroidally compactified.  We shall make
some comments below about the behavior we expect for more
complicated compactification manifolds.  In particular, a more
realistic compactification manifold will determine the form of
the lower dimensional superpotential.

After reducing this action on a torus of compact volume $V$, and
redefining fields to get to the four dimensional Einstein frame,
we recover the four dimensional action (\ref{eq:grav_action}),
with the lower dimensional fields related by
 \begin{eqnarray}
\phi& =& \Phi \\
ds_{10}^2 = G_{MN}dx^M dx^N& =& e^{\frac{1}{2\sqrt{3}}\sigma
+\frac{1}{2}\phi}(e^{-\frac{2}{\sqrt{3}}\sigma}g_{\mu \nu}dx^\mu dx^\nu +
g^c_{mn}dy^m dy^n) \label{eq:ten_metric}\\
\kappa_4^2 &=& \frac{\kappa_{10}^2}{V_6} = \frac{\kappa_{10}^2}{\int d^6y
\sqrt{g^c}} \quad ,
 \end{eqnarray}
where $g^c_{mn}$ is a flat, compact metric on the
six-torus parameterized by $y^m$, whose volume $V_6$ is string-scale.
The 4D dilaton is $\phi$, and
$\sigma$ controls the proper size
of the compact manifold.
Note that the lower-dimensional action is invariant under $SO(2)$
rotations between $\phi$ and $\sigma$. Because of this, there is
some freedom of choice in the definition of the four-dimensional
fields. We have chosen the fields such that the duality relations
for the ten-dimensional fields translate into simple
transformations of the four-dimensional fields. In this paper,
we will be concerned with three varieties of duality
transformation: S-dualities, T-dualities, and ``M-theory limits",
whose action we shall now outline.

The S-duality transformation comes from the stringy strong-weak
coupling S-dualities, i.e. those between the heterotic and type I
strings, or the strong-weak part of the type IIB $SL(2,Z)$
self-duality. These dualities operate on the ten-dimensional  and
string frame metric by,  for example,
 \begin{eqnarray}
G^I_{MN} &=& e^{-\Phi_h}G^h_{MN} \\
\Phi_I &=& -\Phi_h
 \end{eqnarray}
for the type I and heterotic fields, with a similar action occurring
in the type IIB string.  The lower dimensional action thus
inherits a transformation:
\begin{eqnarray}
\label{sdual1}
\phi \rightarrow - \phi \\
\sigma \rightarrow \sigma \quad .
\label{sdual2}
\end{eqnarray}
Note that the S-duality transformations never change the 4
dimensional Einstein frame metric.  We can use these
transformations to choose $\phi \leq 0$, which implies $g_s \leq
1$.

The second type of transformation we shall consider are T-duality
transformations, which for simplicity we shall take to act on the
entire 6-torus:
\begin{eqnarray}
G_c \equiv  e^{\frac{1}{2 \sqrt{3}}\sigma + \frac{1}{2}\phi} g_c
 \rightarrow G_c^{-1}\\
\Phi \rightarrow \Phi - \frac{1}{2}\log \det G_c \quad ,
\end{eqnarray}
which translates into
\begin{eqnarray}
\phi \rightarrow -\frac{1}{2} \phi - \frac{\sqrt{3}}{2}\sigma \\
\sigma \rightarrow - \frac{\sqrt{3}}{2} \phi + \frac{1}{2} \sigma \quad .
\end{eqnarray}
This transformation can be used to choose the combination
$\frac{1}{2 \sqrt{3}}\sigma + \frac{1}{2} \phi \geq 0$, which
causes the compact torus directions to have proper length greater
than or equal to the string length.

The final type of transformation we will consider is the
``M-theory limit".  This limit will be most useful when we are
considering models corresponding to compactifications of the
heterotic $E8 \times E8$ string, although it will also apply to
models derived in the type IIA theory. In solutions where we find
the 10 dimensional string coupling growing large, we expect the
11th dimension of heterotic M-theory to open up.  In the
compactified theory, this should then look like a 5th dimension
opening up.  This is more or less the type of solution that one
deals with in brane world scenarios as well as in the Ekpyrotic
scenario \cite{Ekp1,Ekp2}.
A solution of the four dimensional action maps into a 5
dimensional solution of the Einstein-Hilbert action, coupled to a 5
dimensional scalar field $\sigma_5$, if we identify the 5
dimensional metric and
scalar according to:
\begin{eqnarray}
\label{eq:5limit1}
ds_5^2 &=& e^{-\frac{1}{2} \phi - \frac{1}{2 \sqrt{3}}
\sigma}g_{4 \mu\nu}dx^\mu dx^\nu +
e^{\phi + \frac{1}{\sqrt{3}}\sigma}dz^2 \\
\sigma_5 &=& \frac{\sqrt{3}}{2} \sigma - \frac{1}{2} \phi \quad .
\label{eq:5limit2}
\end{eqnarray}
The 5 dimensional form has been defined so that the coordinate
$z$ is the same one used in the usual $10 \rightarrow 11$
dimensional M-theory decompactification limit, which we recall
takes the form:
\begin{equation}
ds_{11}^2 = e^{-\frac{2 \Phi}{3}} G_{10MN} dx^M dx^N + e^{\frac{4
\Phi}{3}} dz^2 \quad .
\end{equation}
We will also want to pay attention to the 11D metric
corresponding to a given 4D metric, dilaton, and $\sigma$-field:
\begin{equation}
ds_{11}^2 = e^{\frac{1}{2 \sqrt{3}} \sigma - \frac{1}{6}\phi} (
e^{-\frac{2}{\sqrt{3}} \sigma}g_{\mu \nu} dx^\mu dx^\nu + g^c_{mn} dy^m
dy^n) + e^{\frac{4}{3} \phi} dz^2 \quad .
\end{equation}
Note that, while before we chose the compact directions $y^m$ to
be string scale, here we must choose the coordinate $z$ to be of
order  the 11D Planck length $e^{\phi/3} l_s$, for the tower of
Kaluza-Klein states to correctly reproduce the spectrum of
D0-brane bound states.  Requiring both the 11th dimension and the
six compact dimensions to be larger than the 11D Planck length
becomes the requirement that $\phi > 0$ and $\frac{1}{4
\sqrt{3}}\sigma -\frac{5}{12}\phi > 0$, or $\sigma >
\frac{5}{\sqrt{3}}\phi >0$.

Since the dilaton plus gravity effective action of all
perturbative string theories has the same functional form, the
action of dualities on the different effective actions has to be
represented by  field redefinitions which are allowed by the
symmetries of the lagrangian. The duality transformations must
therefore act on the space of solutions by mapping one set of
solutions onto another set.  We will develop these mappings
in the following section.

Note that in general these duality relations are seen in
backgrounds with extended SUSY, such as the toroidal
compactifications we are presently studying.  However, we expect
similar relations to continue to hold at the level of the $N=1$
SUGRA models that we need for a realistic particle phenomenology.  For
example, although it is difficult to describe explicitly for
Calabi-Yau manifolds, our string intuition states that if the
volume modulus shrinks, then various winding states will become
light, and that the appearance of these states can then be
treated by looking at a different effective compactification which has a
large compact volume.
Considering more realistic  $N=1$ SUSY compactifications in our
effective field theory allows non-perturbative superpotentials
for the moduli, whose features we will investigate in section IV.

\section{Outer region solutions without potential}

For regions of our simplified moduli space that have large
compact directions (compared to the string scale) and small
string coupling, we trust the string derivation of the low energy
effective field theory action; this is what we will generically
call the ``safe" region of moduli space If we find ourselves with
$g_s\gg 1$, or a compactification direction $R \ll l_s$, we can
apply an S or T duality to map it to one of the safe regions.
Those parts of moduli space that can be mapped to a safe region
by a duality transformation we will call the outer region of
moduli space.

\begin{figure}[ht]
\begin{center}
\rotatebox{0}{\epsfig{file=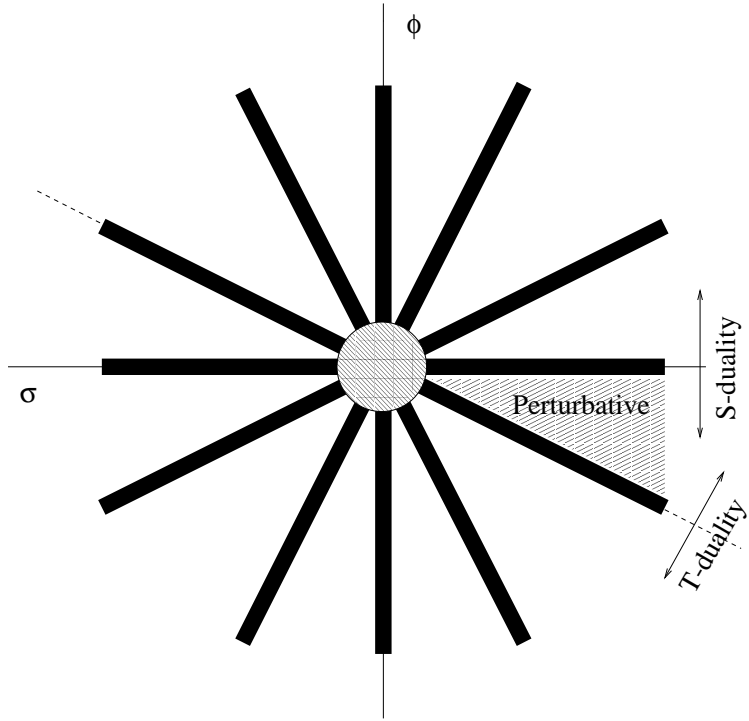,width=10.0cm}}
 \caption{Regions of string moduli space. The shaded circular region is
the ``central" region and the dark rays indicate regions in which
one of the moduli is perturbative while the other is not. The
region marked ``perturbative" is the fully perturbative region and
the white wedges between the rays are its images under duality.}
\label{dual1}
\end{center}
\end{figure}

Figure~1 shows the perturbative region and its images under
different dualities for backgrounds that exhibit S-duality.
The ray along
the $\phi$ axis depicts the
region in which $\sigma$ is approximately fixed such that the
compact volume is of order the string volume and the coupling is
perturbative for {\em some} background. The ray along the $\sigma$ axis
depicts the region in which $\phi$ is approximately fixed such
that the string coupling is of order one  and the compact volume
is larger than the string volume for some background. The other
rays can be thought of as their images under duality.
Note that along the $\sigma$ axis, for example, we expect to be able
to use
four-dimensional $N=1$ SUGRA with {\em one} dynamical field $\phi$ to
study cosmology. However, we do not have a way to derive the effective
action in a trustworthy manner from M-theory. Thus to study
cosmology in this region of moduli space we choose to parameterize
it by a chiral superfield of $D=4$, $N=1$ SUGRA.

The regions with both $g_s \sim 1$ and $R \sim l_s$, which cannot
be mapped to a safe region for either modulus by applying a
duality transformation is called the ``inner" (or ``central")
region, and it is where we expect the general principles of
string universality to apply \cite{bda1}. Again, we expect to be
able to use $N=1$ SUGRA in studying cosmology here, although we
lack a way to derive the effective action systematically from
M-theory. Thus to study cosmology in the central region of moduli
space we generally parameterize it  by several chiral superfields
of D=4, $N=1$ SUGRA. We assume, along the lines of \cite{bda2},
that they are all stabilized at the string scale by stringy
non-perturbative (SNP) effects which allow a continuously
adjustable constant in the superpotential. SUSY is broken at an
intermediate scale by field theoretic effects that shift the
stabilized moduli only by a small amount from their unbroken
minima. The cosmological constant can be made to vanish after
SUSY breaking by the adjustable constant.

In the rest of this section, we explicitly exhibit the duality
relations that can be used to map ``unsafe" region solutions of
the outer region to ``safe" regions, which allows us to get a
grasp on the cosmological behavior of any model in the outer
region.  With this we will be able to argue that realistic
slow-roll inflationary cosmology must, within our general
assumptions, necessarily occur in the (relatively quite small)
central region of our simplified moduli space.

An important caveat to mention is that our knowledge of the
behavior within moduli space is more complete in those models
exhibiting S-duality. In particular, the picture of calculable
regions within moduli space for models with M-limits is as shown
Figure~2.  There is a large region where the dilaton is large
enough that we need to use the 11D theory, but also in which the
6 compact dimensions are smaller than the 11D Planck length.
Lacking any analog of T-duality for membrane states, we end up
with a larger unsafe region than just the central region of
interest. Perhaps this region can still be parametrized by a 5D
SUGRA, so some general conclusions might still be made. In any
case, we will later see that membrane instantons lead to an
instability analogous to the Dine-Seiberg instability, and thus
that we are once again forced to consider the central region once
we allow for scalar potentials.  It should be noted,
however, that knowledge of the general behavior of potentials for
M-theory limits is not as complete as it is for the perturbative
string backgrounds.

\begin{figure}[ht]
\begin{center}
\rotatebox{0}{\epsfig{file=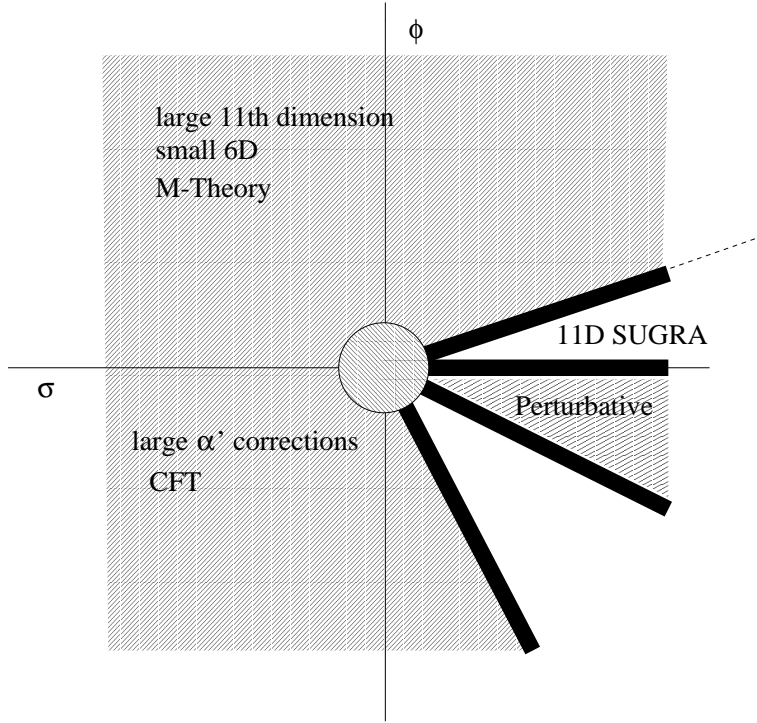,width=10.0cm}}
 \caption{Regions of M-theory moduli space. The shaded circular region is
the ``central" region and the dark rays indicate regions in which
one of the moduli is perturbative while the other is not. The
region marked ``perturbative" is the fully perturbative region.
Other regions are marked accordingly: the wedge where 11D SUGRA
is valid, the shaded region in which the full quantum M-theory
needs to be used, and the region in which classical string theory is
valid and $\alpha'$ corrections are large. The unmarked wedge is T-dual to 
the perturbative region}
\label{dual2}
\end{center}
\end{figure}

As explained previously, in terms of the SUGRA fields of the low
energy theory the duality transformations correspond to trivial
field redefinitions, which are allowed by the symmetries of the
lagrangian.  However, we should always keep in mind that the model
is derived from a more fundamental string theory. Because of the
duality relations, any solution in the outer region can be
derived in a trustworthy manner from some perturbative string
theory, within which solutions are well known and offer very few
possibilities for surprises.

We begin by investigating models with 4 dimensional flat
cosmologies, and a time dependent dilaton, which will capture the
essential physics of any outer region cosmologies without
potentials. We will use the Friedmann-Robertson-Walker (FRW)
 ansatz for
the non-compact space:
\begin{eqnarray}
ds_4^2 &=& -dt^2 + a^2(t)dx^i dx^i \\
\phi &=& \phi(t) \quad .
\end{eqnarray}
With this ansatz for the fields, and with our fields normalized as in
eq.(\ref{eq:grav_action})
we find the equations of motion:
\begin{eqnarray}
H^2 &=& \frac{1}{6} \rho \\
\dot{H} &=& -\frac{1}{4}(\rho +p) \\
\ddot{\phi}+3 H\dot{\phi}&=&0 \quad ,
\label{eq:eom1mod}
\end{eqnarray}
where
\begin{eqnarray}
H &=& \dot{a}/a\\
\rho&=&p=\frac{1}{2} \dot{\phi}^2 \quad .
\end{eqnarray}
The cosmological solutions to these equations are simple and well
known,
\begin{eqnarray}
\label{eq:sol1mod}
a(t) &=& a(1)|t|^{1/3} \\
\phi(t) &=& \phi(1) \pm \frac{2}{\sqrt{3}} \ln |t| \quad ,
\label{eq:sol2mod}
\end{eqnarray}
where we have placed the inevitable singularity in the solution
at $t=0$ for convenience. Of course, the solutions cannot be
trusted in regions where the string coupling is of order unity
and the curvature is high.

These well-known solutions naturally break up into  four
different cases, depending on the sign chosen in the solution for
$\phi$, and on whether the singularity in $t$ lies to the future
or to the past. In the Pre-Big Bang (PBB) scenario, and in the
Ekpyrotic scenario, a  transition between the different branches
is envisioned in a single cosmology of the same string theory.
We, on the other hand, consider duality relations that map
solutions (whether they include a transition between the branches
or not) onto other well controlled solutions of (in general) {\em another}
string theory.

\begin{figure}[ht]
\begin{center}
\rotatebox{0}{\epsfig{file=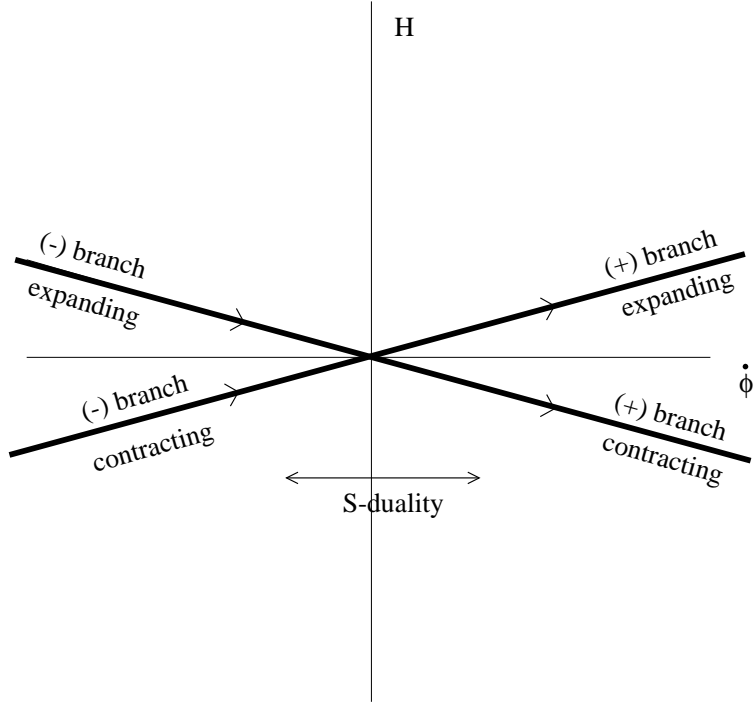,width=10.0cm}}
\caption{Action of S-duality on cosmological solutions of
different string theories. The solutions have the same form in
each theory, and the arrow labeled ``S-duality" indicates how they
are related. For example, the contracting  $(+)$ solution in one
theory is mapped onto the contracting $(-)$ solution of the
S-dual theory.}
 \label{dual3}
\end{center}
\end{figure}

In this simple first case, we can apply either S-duality
transformations or M-theory limits.  The S-duality transformation
will apply for a model that is derived from type IIB, type I, or
Heterotic $SO(32)$ string theories, and simply reverses the sign
of the dilaton according to eqs.(\ref{sdual1}),(\ref{sdual2}),
leaving the metric unchanged.  Thus it simply exchanges the `$+$'
and `$-$' branches of the solutions of the two dual theories. The
``pool" of solutions to choose from in any of the theories is the
same pool for all theories. For models derived from the Heterotic
$E8 \times E8$ or type IIA theories, our solutions map into
5-dimensional solutions, using eqs.~(\ref{eq:5limit1}),
(\ref{eq:5limit2})
\begin{eqnarray}
 ds_5^2 &=& e^{-\frac{\phi(1)}{2}} |t|^{\pm \frac{1}{\sqrt{3}}}(-dt^2 +
a^2(1) |t|^{\frac{2}{3}}dx^i dx^i) + e^{\phi(1)}|t|^{\pm
\frac{2}{\sqrt{3}}} dz^2 \\
\sigma_5 &=& -\frac{1}{2} \phi(1) \mp \frac{1}{\sqrt{3}} \ln|t| \quad ,
\end{eqnarray}
where $\sigma_5$ is a massless scalar in 5D.
One can easily check that this is a solution of the 5-dimensional
equations of motion.

We would like to point out that the 5D solution space is isomorphic
to the 4D solution space, and therefore 4D solutions can be
mapped onto 5D solutions and vice-versa. Properties of solutions
can be studied in either the 4D or the 5D setup.  The 5D solution is
relevant for compactifications that have a strong coupling limit described
in terms of 11D  SUGRA.

We make the following almost trivial observations:\\
$(i)$\ Solutions (\ref{eq:sol1mod}), (\ref{eq:sol2mod}) do not describe
slow-roll inflation at any time, since they obey $\dot{H} \sim H^2$.
They {\em can} describe a period of accelerated contraction as in the PBB
scenario,
which could be interpreted as ``fast-roll" inflation.\\
$(ii)$\
Obviously, the moduli are not stabilized wherever equations
(\ref{eq:sol1mod}) and
 (\ref{eq:sol2mod}) are valid.\\
$(iii)$\ As the solution approaches the central region it ceases
to be valid and new interaction terms in the effective action are
expected to kick in.\\
These conclusions are valid for 4D
solutions as well as for 5D solutions, where they are interpreted
as the motion of branes in a 5D universe.

We can now relax our restrictions somewhat, and consider the case
with two moduli.  We will consider one modulus, $\phi$,
controlling the coupling, while the other, $\sigma$, controls the
size of the compact manifold through our  ansatz
(\ref{eq:ten_metric}) for the 10D metric. We now have an
additional equation of motion,
\begin{equation}
\ddot{\sigma}+ 3 H\dot{\sigma}=0 \quad ,
\label{eq:eom2mod}
\end{equation}
and now
\begin{equation}
\rho=p=\frac{1}{2} \dot{\phi}^2 + \frac{1}{2} \dot{\sigma}^2 \quad .
\end{equation}

The 4D solutions now take the more
general form
\begin{eqnarray}
\label{eq:sol2moda}
a(t) &=& a(1)|t|^{\frac{1}{3}} \\
\label{eq:sol2modphi}
\phi(t) &=& \phi(1) + p_\phi ln |t| \\
\label{eq:sol2modsigma}
 \sigma(t) &=& \sigma(1) + p_\sigma ln|t| \quad ,
\end{eqnarray}
with the algebraic constraint
\begin{equation}
p_\phi^2 + p_\sigma^2 = \frac{4}{3} \quad .
\end{equation}
For weak coupling and low string-curvature solutions, we require
$\phi < 0$ and $\frac{1}{2 \sqrt{3}} \sigma + \frac{1}{2} \phi >
0$, and we find ourselves in the central region when
$\sigma,\phi\sim 0$. Note that, apart from some simple mixing
between $\phi$ and $\sigma$ as sources, the dynamics is identical
to the single modulus case, in that there will be solutions
looking like either the `$+$' or `$-$' branches considered
previously, and that they can be studied equivalently in their 4D
or 5D versions, as we show below.

As in the single field case, we note that solutions
(\ref{eq:sol2moda}), (\ref{eq:sol2modphi}), and
(\ref{eq:sol2modsigma}) do not describe slow-roll inflation at
any time, but they can describe a period of accelerated
contraction as in the PBB scenario, that  moduli are not
stabilized while (\ref{eq:sol2moda}), (\ref{eq:sol2modphi}), and
(\ref{eq:sol2modsigma}) are valid, and that as the solution
approaches the central region it ceases to be valid and new
interaction terms in the effective action are expected to kick in.

A detailed study of such solutions and their generalizations has
been conducted for many specific cases: PBB, heterotic
M-theory, Ekpyrotic scenario, etc. Our aim here is to show that
the classification of solutions in the outer region is complete
and that there are no new features coming from any particular
microscopic dynamics that one may wish to invoke in the outer
region.

Now that we have a two moduli set-up, we can also look at the
action of T-duality on the solutions. The only real issue is to
pay attention to when a given solution is in the outer region,
and when it is in the central region.

Let's first look at the case of models derived from type IIB,
type I, or Heterotic $SO(32)$ strings. As argued previously on
general grounds, since the action for these theories is of the
same form, and therefore the spaces of solutions of the theories
are isomorphic, dualities must have a representation on the space
of solutions. For this particular case the 4D
solutions are mapped to other 4D solutions under S-duality by
\begin{eqnarray}
\phi(1) \rightarrow - \phi(1)\\
p_\phi \rightarrow - p_\phi \quad ,
\end{eqnarray}
and $\sigma$ is unchanged.
Under T-duality the solutions undergo a more complicated transformation
\begin{eqnarray}
p_\phi \rightarrow -\frac{1}{2} p_\phi  - \frac{\sqrt{3}}{2} p_\sigma \\
p_\sigma \rightarrow - \frac{\sqrt{3}}{2} p_\phi + \frac{1}{2}
p_\sigma \quad ,
\end{eqnarray}
with an identical action holding on the integration constants $\phi(1)$
and $\sigma(1)$.  These are clearly still solutions, since
\begin{equation}
(-\frac{1}{2} p_\phi  - \frac{\sqrt{3}}{2} p_\sigma)^2 +
(- \frac{\sqrt{3}}{2} p_\phi + \frac{1}{2} p_\sigma)^2 = p_\phi^2 +
p_\sigma^2 \quad .
\end{equation}

Notice that despite the very similar form, this transformation is
very different from the one induced by ``Scale-factor-duality"
\cite{SFD} which is a symmetry of the equations of motion for
cosmological solutions under which solutions are mapped onto
other solutions of the same string theory, while dualities map
solutions of one string theory onto solutions of the {\em dual}
string theory.

The two moduli solutions that we have found can be explicitly
mapped to 5D, 10D, and 11D solutions, as we show below and in the
appendix. It is the 10D form of the solution that determines
whether we trust string perturbation theory; when the string
frame curvature is small in string units, the compact directions
are large compared to the string scale, and the string coupling
is small, we can follow the evolution with confidence. The 5D
solution is of interest whenever all relevant length scales are
much larger than the 5D Planck scale, giving us a 5D
interpretation of cosmology, and the 11D solution is of interest
when it is trustworthy in 11D SUGRA.

We will discuss explicitly the 5D case appropriate for studying
braneworld models, and relegate the other examples to Appendix A.
For the 5D case we have Einstein gravity coupled to one scalar
$\sigma_5$ as in eqs.(\ref{eq:5limit1},\ref{eq:5limit2}). In this
case the map into 5D solutions is the following:
\begin{eqnarray}
 \label{solmap5d}
ds_{5}^2 &=& C_5^2 |t|^{-\frac{1}{2 \sqrt{3}} p_\sigma -
\frac{1}{2}p_\phi} (- dt^2 + a^2(1)|t|^{\frac{2}{3}} dx^i dx^i) +
C_5^{-4}
|t|^{\frac{1}{\sqrt{3}} p_\sigma + p_\phi} dz^2\\
\sigma_5 &=& \frac{\sqrt{3}}{2} \sigma(1) - \frac{1}{2} \phi(1) +
(\frac{\sqrt{3}}{2} p_\sigma - \frac{1}{2} p_\phi) \ln{|t|} \\
C_5 &=& e^{-\frac{1}{4 \sqrt{3}} \sigma(1) - \frac{1}{4} \phi(1)} \quad .
 \end{eqnarray}
These become, after rescaling the time variable by $dt_{5} = C_5
|t|^{-\frac{1}{4 \sqrt{3}} p_\sigma - \frac{1}{4} p_\phi} dt$,
\begin{eqnarray}
ds_5^2 &=& -dt_5^2 + a_5^2(1)|t_5|^{2p_{5a}} dx^i dx^i +
c_5^2(1)|t_5|^{2p_{5c}} dz^2 \\
\sigma_5 &=& \sigma_5(1) + p_{5\sigma}\ln{|t_5|} \quad .
\end{eqnarray}
More details can be found in the appendix.

To summarize the behavior of the outer region solutions, we note
that such solutions do not describe slow-roll inflation; when
they are valid the moduli are not stabilized and they become
invalid as they approach the central region. This holds for 4D
solutions and for their 5D counterparts. Our conclusions are the
following

\begin{enumerate}
\item
It does not seem to be possible to construct a viable cosmology
which evolves entirely in the outer region.
\item
Slow-roll inflation is likely to be obtained only when moduli are
in the central region.

\end{enumerate}
In the next section we will show that these conclusions are not
likely to be altered when potentials for the moduli are included.

\section{Cosmology with moduli potentials}

The moduli potential is an important ingredient in any string
theoretic cosmological scenario, and it is particularly important
when looking for scenarios that involve an epoch of slow-roll
inflation. We will treat this in the same spirit as
the rest of the paper:
we look for moduli potentials that are either derived from string
theory, in cases where the derivation is trustworthy, or that are
constrained by general arguments and symmetries when such a
derivation is not available. In addition, we consider constraints
coming from the requirement of moduli stabilization in a minimum
with an acceptably small cosmological constant, in which moduli
can end up after the inflationary epoch.  This turns out to
impose rather severe restrictions on models. We will continue to consider
 theories that are imbedded in $N=1$
SUGRA.

In the context of weakly coupled heterotic string theory it is
well known that the potentials that are generated for the dilaton
and the compactification moduli are typically of the runaway form
\cite{DS} so that the theory prefers to go to the zero coupling
and decompactification limit. Similar problems are encountered in
theories in which SUSY is broken completely by string theoretic
effects, such as the Scherk-Schwarz mechanism where the
compactification explicitly breaks SUSY. In addition, the
generated potentials are steep \cite{BS}, and consequently have a
problem providing enough slow-roll inflation. We will use
dualities to extend these arguments to all corners of moduli
space. Additionally, we will include the contributions of brane
instantons. In the next section models of brane annihilation,
tachyon condensation, and Ekpyrosis will be discussed and the
problems with such scenarios will be highlighted.

We focus on non-perturbative potentials, since it is well-known
that perturbative potentials have problems in satisfying the
generic phenomenological requirements as above. Non-perturbative
sources for moduli potentials originate from field theoretic
effects and stringy non-perturbative effects (SNP). Field
theoretic effects such as hidden sector gaugino condensation
\cite{gaugino1,gaugino2}
have
been discussed previously very extensively
(see, for example, \cite{Nilles:1998uy}), so we will not
repeat
the discussion here. The SNP that we will consider originate from
brane-instanton effects.

General arguments based on Peccei-Quinn symmetries, and how they
break due to non-perturbative effects, show that the
superpotential must be a sum of exponentials in the moduli. These
exponentials can be generated by either stringy or field theoretic
non-perturbative effects. A constant term in the superpotential is
allowed by the symmetries but has no natural mechanism for its
generation.

In the last few years there have been attempts to stabilize
moduli using various fluxes in the internal space (see for
example \cite{gkp}). However all these arguments ignore the back
reaction of the flux on the internal space and it appears that
when this is taken into account the derivation of the potential for the moduli
becomes invalid. A detailed discussion of the problems associated with 
potentials from fluxes is currently under preparation by one of the authors
(SdA). In any case, even according to these discussions the
volume modulus is not stabilized, so that one would still be
forced to consider SNP effects. Also these potentials are
qualitatively similar to those generated by gaugino condensation
(i.e. the so-called race-track models 
\cite{kras,deCarlos:1992da,Dine:1999dx}) and hence will have the
same cosmological problems as those highlighted in \cite{BS}.

To some extent the results of this section are contained
explicitly or implicitly in the existing literature, but we will
state them explicitly to substantiate our arguments.

\subsection{Non-perturbative moduli potentials}

Even in the absence of SUSY breaking one would expect SNP effects
coming from the various brane instantons\cite{bda1,ovrut,moore}.
Our goal is to determine their influence on cosmological
solutions. Brane instantons are Euclidean configurations that
are obtained by wrapping the extended directions of a brane,
including its Euclidean time direction, around some cycle in the
compact space. Their action is the product of the brane tension and
the volume of the cycle around which its world volume is wrapped.
Since under dualities  BPS branes transform into BPS branes,
brane instanton SNPs in one string theory (or 11D SUGRA)
transform into SNPs in the other theory.

For example, in the case of the M-limit transformation between the
heterotic
\( E_{8}\times E_{8} \) (HE) theory and its strong coupling
version the Horava-Witten (HW) theory, the non-perturbative
effects due to the fundamental string and the NS fivebrane
instantons tend to give a runaway dilaton potential which makes
the dilaton roll down to weak coupling, whereas in the HW theory
the corresponding dual effect originates from  the M two-brane
and M five-brane instantons which tend to push the theory
towards infinite radius for the eleventh dimension, i.e. to strong
coupling.  In terms of the dilaton, these effects are competing and it is
thus plausible to have it stabilized around zero, where neither
perturbative heterotic strings nor 11D supergravity can describe the
physics.

Note that an additional effect of the M-brane instantons for the
two-modulus case is to drive the 6 compact directions $y_m$ to
large radius as well.  However, as illustrated in Figure 2, there is a
larger region of moduli space for ``M-limit" theories that are not
accessible to direct calculation.  We are thus unable, without a more
detailed treatment of the superpotential, to exhibit the competing
effects that would stabilize a second modulus like the $\sigma$ field.

As in the previous sections we focus for simplicity on toroidally
compactified theories and we therefore have two moduli: $\phi$,
which controls the string coupling,
\begin{equation}
\label{coupl1}
 g=e^{\phi} \quad ,
\end{equation}
 and $\sigma$, which controls the compactification radius,
\begin{equation}
 \label{rad1}
 R/l_S=e^{\frac{1}{4\sqrt{3}}\sigma+\frac{1}{4}\phi} \quad .
 \end{equation}
For the M-limit case, when an extra dimension opens up the radius
of the 11th dimension obeys
 \begin{equation}
 \label{rad11}
 \rho/l_{11} =
e^{\frac{2}{3}\phi} \quad ,
\end{equation}
 and the other six compact dimensions have
radius
 \begin{equation}
 \label{rad2}
 R/l_{11}=e^{\frac{1}{4\sqrt{3}}\sigma -
\frac{5}{12}\phi} \quad .
\end{equation}

In accordance with the general arguments about the dependence of
SNP  on moduli, it was found that SNP effects depend
exponentially both on the compact volume $V_6$ and on either $1/g$ or
$1/g^2$. In the outer region of moduli space dominant SNP come
from two or sometimes three BPS branes, whose instanton actions
have the weakest dependence on the string coupling $g$, and
compactification radius $R$, or their product. A list of such BPS
branes and their instanton actions is given in Table~\ref{tbI}.
The notation is taken from \cite{bda1}.

\begin{table}[ht]
    \begin{center}
    \begin{tabular}{||c|c|c||c|c||}
    \hline \hline
     & \ \ \ Type I \ \ \  &   & \ \ \ HO  \ \ \  &   \\
    \hline\hline
     &  brane  &\ \ \  action \ \ \ & brane & \ \ \ action \ \ \ \\
   [.5ex]  \hline
     & $ D5$ & $
    \frac{1}{g_I} \left(\frac{R}{l_{I}}\right)^6  $
    & $ F5 $ &  \
    $\frac{1}{g_{HO}^2} \left(\frac{R}{l_{HO}}\right)^6 $ \  \\
    [.5ex] \hline \hline
     & $ D1 $  &
    $ \frac{1}{g_I} \left(\frac{R}{l_{I}}\right)^2  $
    & $ F1$  & $ \left(\frac{R}{l_{HO}}\right)^2 $ \\
       [.5ex] \hline\hline
  \hline \hline
     & \ \ \ HE \ \ \  &   & \ \ \ HW  \ \ \  &   \\
    \hline\hline
     &  brane  &\ \ \  action \ \ \ & brane & \ \ \ action \ \ \ \\
   [.6ex]  \hline
 \  \ & $ F5 $ & $ \frac{1}{g_{HE}^2} \left(\frac{R}{l_{HE}}\right)^6$
    &\ $ M5 $ transverse \ & $ \left(\frac{R}{l_{11}}\right)^6$ \  \\
    [.6ex] \hline \hline
   \  \ &\ $ F1 $\ & $ \left(\frac{R}{l_{HE}}\right)^2  $
    &\ $ M2 $ longitudinal\  & $
    \left(\frac{R}{l_{11}}\right)^2 \frac{\rho}{l_{11}} $ \\
       [.6ex] \hline\hline
  \hline \hline
     & \ \ \ IIA \ \ \  &   & \ \ \ MS1  \ \ \  &   \\
    \hline\hline
     &  brane  &\ \ \  action \ \ \ & brane & \ \ \ action \ \ \ \\
   [.6ex]  \hline
    & $ D0 $ & $ \frac{1}{g_{IIA}} \frac{R}{l_{IIA}}  $
    & $ KK $ graviton & $ \frac{R}{\rho}$ \  \\
    [.6ex] \hline \hline
      &\ $ F1 $ & $ \left(\frac{R}{l_{IIA}}\right)^2  $
    & $ M2 $ longitudinal  & $
    \left(\frac{R}{l_{11}}\right)^2 \frac{\rho}{l_{11}} $\ \\
       [.6ex] \hline\hline
\hline \hline
     & \ \ \ IIA \ \ \  &  & \ \ \ IIB  \ \ \  &   \\
    \hline\hline
     &  brane  &\ \ \  action \ \ \ & brane & \ \ \ action \ \ \ \\
   [.6ex]  \hline
  L & $ D0 $ & $ \frac{1}{g_{IIA}} \frac{R_{IIA}}{l_{II}}  $
    &  $D\!-\!1$& $ \frac{1}{g_{IIB}}$ \  \\
    [.6ex] \hline \hline
    T & $ D0 $ & $ \frac{1}{g_{IIA}} \frac{R}{l_{II}}$
    &  $D1$ & $ \frac{R R_{IIB}}{g_{IIB}l_{II}^2}$ \  \\
    [.6ex] \hline \hline
   L & $ F1$ & $ \frac{RR_{IIA}}{l_{II}^2}$
    & {$KK$ MM} & $ \frac{R}{R_{IIB}}$ \  \\
    [.6ex] \hline \hline
    T & $ F1 $ & $ \frac{R^2}{l_{II}^2}$
    &  {$F1$ } & $ \frac{R^2}{l_{II}^2}$ \  \\
    [.6ex] \hline \hline
  \end{tabular}
\caption{Dominant BPS branes and their instanton actions taken
from \cite{bda1}.}
  \label{tbI}
\end{center}
  \end{table}

In the table, $Dn$ refers to a D-brane instanton with an
$n+1$-dimensional world-volume in the appropriate theory.
Similarly $F1$ and $F5$ refer to instantons for the fundamental
string or NS 5-brane wrapping 2D or 6D volumes, respectively.
Longitudinal and transverse refer to whether or not the brane
wraps the M-limit direction. In the portion of the table with
instantons for the type IIA and type IIB branes, the two
backgrounds are related by T-duality in a preferred direction.
The entries marked $L$ and $T$ refer to wrappings that are
longitudinal or transverse with respect to this preferred
coordinate. $KK$ MM stands for Kaluza-Klein momentum mode. The
size of a direction that is not T-dualized is denoted $R$. Unlike
the previous cases of S-duality, we have to make a distinction
between different radii, since even if their sizes were equal to
begin with, the T-duality changes them.

To determine cosmology in the outer region of moduli space it is
enough to consider only a few contributions. The (super)potentials
induced by brane instantons, in the approximation that we will
use, ignoring the prefactor and questions related to fermion zero
modes, are simply given by $V\sim e^{- {\rm action}}$.

Note that in terms of our scalars $\phi$ and $\sigma$, the brane
instanton potentials are exponentials of exponentials in
the moduli fields,  for
example, $V\sim e^{-e^{-\phi}}$.
Duality acts on cosmological equations of motion
with a potential induced by SNP in one background by mapping into the
equations of another theory, as in the case without potentials.
Therefore solutions in one theory have to be mapped onto
solutions of the other theory. The exact form of the
solutions cannot be obtained analytically, but their general
properties are quite similar to the no-potential case.

As an example of this mapping, we will focus on the S-duality
between the type I and HO theories.  The type I theory has $D1$
and $D5$-brane instantons, wrapped on the compact dimensions,
that when expressed in terms of our 4D scalars take the form:
 \begin{eqnarray}
 \label{actd5}
D5&:& V \sim e^{\hbox{$-\left[
{e^{\frac{\sqrt3}{2}\sigma +\frac{1}{2}\phi}}\right]$}}\\
 \label{actd1}
 D1&:& V \sim e^{\hbox{$-\left[{e^{\frac{1}{2
\sqrt{3}}\sigma - \frac{1}{2}\phi}}\right]$}} \quad .
\end{eqnarray}

The actions for the $D5$-brane and $D1$-brane instantons were taken
from the appropriate entries in Table I, and evaluated using
eq.(\ref{coupl1}) and eq.(\ref{rad1}). Meanwhile, an HO background
with $F5$ and $F1$ instantons wrapping compact dimensions would
give the contributions:
\begin{eqnarray}
 \label{actf5}
F5&:& V \sim e^{\hbox{$-\left[
{e^{\frac{\sqrt3}{2}\sigma - \frac{1}{2}\phi}}\right]$}}\\
 \label{actf1}
F1&:& V \sim e^{\hbox{$-\left[
 {e^{\frac{1}{2 \sqrt{3}}\sigma + \frac{1}{2}\phi}}\right]$}} \quad ,
\end{eqnarray}
which were evaluated using eqs.(\ref{coupl1}),(\ref{rad1}).

Of course, it is no surprise that the action of S-duality ($\phi
\rightarrow - \phi$)
exchanges the type I and HO instantons.  T-duality also offers no
real surprises, but does bring other instantons into play.  In
particular, the action of T-duality on all compact radii maps the
$D5$ instanton to a $D(-1)$ instanton with action $e^{-\phi}$ and
the $D1$ instanton to a $D3$ instanton with action $V \sim e^{-
1/g_I (R/l_I)^4}$, which using eqs.(\ref{coupl1}), and
(\ref{rad1}) can be shown to be equal to $V \sim
e^{-\hbox{$ e^{\frac{1}{\sqrt{3}}\sigma}$} }$.

On the HO side, the $F5$ instanton is invariant under T-duality,
while the $F1$ instanton becomes the contribution of a
Kaluza-Klein momentum mode with action $\sim m^2 \sim
\frac{1}{R^2} \sim e^{-\frac{1}{2} \phi - \frac{1}{2
\sqrt{3}}\sigma}$. This mapping can be extended to the other
cases, such as the M-limits. However, since one of the main
pieces of evidence for duality comes from the spectrum of BPS
states, and we are looking at contributions from BPS instantons,
it is obvious that this mapping will work.

\subsection{Slow-roll inflation}

We would like to discuss the possibility of obtaining slow-roll
inflation in the outer region of moduli space in the presence of
non-perturbative potentials. Since it is not possible to find
analytic solutions for the moduli equations of motion in the
presence of non-perturbative moduli potentials, it is more
convenient to do the analysis with solvable exponential
potentials, and then to argue using these results about the generic
behavior of solutions with non-perturbative potentials. Here we
do not need to consider explicitly the influence of $N=1$ SUGRA
and of extra dimensions. It is enough to consider four dimensional
bosonic actions.

Consider adding to the moduli action (\ref{eq:grav_action})
an exponential potential
term whose generic form is $V(\phi,\sigma)=A
e^{\alpha\sigma+\beta\phi}$,
\begin{equation}
S_4=\int d^{4} x \sqrt{-g}\left\{R- \frac{1}{2}
(\partial\phi)^2-\frac{1}{2}(\partial\sigma)^2-
A e^{\alpha\sigma+\beta\phi}\right\} \quad .
\end{equation}
The resulting equations of motion are the following,
\begin{eqnarray}
 \label{exppot1a}
H^2 &=& \frac{1}{6} \rho \\
 \label{exppot1b}
\dot H &=& - \frac{1}{4} (\rho+p) \\
 \label{exppot1c}
\ddot{\phi} &+& 3 H \dot\phi +
\frac{\delta V}{\delta\phi} =0 \\
\label{exppot1d} \ddot{\sigma} &+& 3 H \dot\sigma + \frac{\delta
V}{\delta\sigma} =0 \quad ,
\end{eqnarray}
where
\begin{eqnarray}
\rho &=& \frac{1}{2} \dot\phi^2+\frac{1}{2} \dot\sigma^2 +
V(\phi,\sigma) \\
\rho +p &=&  \dot\phi^2+ \dot\sigma^2 \quad .
 \label{exppot2}
\end{eqnarray}
Only three of the equations (\ref{exppot1a}-\ref{exppot1d}) are
independent. For example, eq.(\ref{exppot1d}) can be obtained by
taking a time derivative of eq.(\ref{exppot1a}) and substituting
in eqs.(\ref{exppot1b}),(\ref{exppot1c}).

We may change variables to
$\psi_1=\frac{\alpha\sigma+\beta\phi}{\sqrt{\alpha^2+\beta^2}}$
and
$\psi_2=\frac{-\beta\sigma+\alpha\phi}{\sqrt{\alpha^2+\beta^2}}$.
Then $\dot\psi_1^2+\dot\psi_2^2= \dot\phi^2+\dot\sigma^2$, and
the potential $V$ is only a function of $\psi_1$, $V=A
e^{\sqrt{\alpha^2+\beta^2} \psi_1}$.

Assuming that solutions take the form
\begin{eqnarray}
a(t)&=&a(1) |t|^{p_a}\\
\psi_1(t)&=&\psi_1(1)+p_1\ln|t| \\
\psi_2(t)&=&\psi_2(1)+p_2\ln|t| \quad ,
\label{solexppot}
\end{eqnarray}
and denoting $\gamma=\sqrt{\alpha^2+\beta^2}$, we find the
following set of algebraic equations,
\begin{eqnarray}
\label{exppar1a}
\gamma p_1 &=& -2 \\
\label{exppar1b}
p_a^2&=& \frac{1}{6}\left(
 \frac{1}{2} p_1^2+\frac{1}{2} p_2^2 + A e^{\gamma\psi_1(1)} \right) \\
 p_a &=& \frac{1}{2}\left(
 \frac{1}{2} p_1^2+\frac{1}{2} p_2^2 \right) \\
\label{exppar1c}
-p_1&+&3 p_a p_1 +A\gamma e^{\gamma\psi_1(1)}=0 \\
\label{exppar1d} -p_2&+&3 p_a p_2 =0 \quad .
\end{eqnarray}

The solutions have similar forms to those without a potential.

{} From eq.(\ref{exppar1d}) we obtain $p_2(1-3 p_a)=0$,
so either $p_2=0$ or $1-3 p_a=0$,
but from eq.(\ref{exppar1c}) we obtain
$p_1(1-3 p_a)= A\gamma e^{\gamma\psi_1(1)}$, so
$1-3 p_a=0$ can only be a solution if $A=0$ or $\gamma=0$,
which recovers the case with no potential. The only choice
then is to have $p_2=0$, that is a constant $\psi_2$ which in
effect reduces the problem to the single field case.

Setting $p_2=0$ in eqs.(\ref{exppar1a}-\ref{exppar1d}) we obtain
\begin{eqnarray}
\label{exppar1e}
\gamma p_1 &=& -2 \\
\label{exppar1f}
p_a^2&=& \frac{1}{6}\left(
 \frac{1}{2} p_1^2+ A e^{\gamma\psi_1(1)} \right) \\
 \label{exppar1g}
 p_a &=& \frac{p_1^2 }{4} \\
\label{exppar1h} -p_1&+&3 p_a p_1 +A\gamma e^{\gamma\psi_1(1)}=0 \quad ,
\end{eqnarray}
where only three equations are independent, so we may ignore
eq.(\ref{exppar1h}). We may also substitute eq.(\ref{exppar1e})
into eq.(\ref{exppar1g}) and obtain
\begin{eqnarray}
 p_a &=& \frac{1 }{\gamma^2 } \quad .
\end{eqnarray}
For a solution to exist for large values of $\gamma$, the
prefactor $A$ needs to be negative.
For inflation, we require $\dot{a}$ and $\ddot{a}$ to be
greater than zero, which implies $\gamma^2<1$. Thus we require a
potential that is flat enough.

The ansatz that we have used does not give the most general set of
solutions to the equations of motion.  This is given in
\cite{coprev}. However, the solutions that can be found using it
are enough for our purposes, since we wish to determine the
dependence of the character of the solution on the steepness of
the potential, and in particular we wish to show that steep
potentials cannot support slow-roll inflation.

The brane instanton potentials are exponentials of exponentials,
as has been illustrated in our discussion of the type I and HO
duality mappings. From this, we in general find $V'/V\sim
1/g_s, 1/g_s^2$, $V'/V\sim R/l_s$, or some
similar combination,
which is large in the outer region of moduli space (where
$\frac{1}{g_s}, \frac{R}{l_s}\gg 1$), implying that it is difficult to get
a potential that is flat enough for inflation. Indeed, here we may
use the old trick of \cite{BS} to compare exponential potentials
to exponentials of exponentials. For a given non-perturbative
potential $V$, one finds an exponential potential $U$ which
matches its value and the value of its derivative $V'$ for a
given value of the field $U(\phi_0)=V(\phi_0)$,
$V'(\phi_0)=U'(\phi_0)$. As we have seen, it is easy to determine
whether or not an exponential potential will lead to slow-roll
inflation. In the outer region $U$ will be steep, and therefore
cannot support slow-roll inflation. Now, we know that the
exponential of an exponential $V$ is even steeper than the
exponential potential $U$, and so, based on our general
arguments, will also not support slow-roll inflation.

What we have learned from the previous exercise is that the rate
of scale factor expansion is determined by competition in the
conversion of potential energy into kinetic energy, in which  the
field whose potential is the steepest wins. In 4D SUGRA obtained
in the perturbative region we know that the Kahler potentials of
the coupling and volume moduli are $-\ln (S+\bar S)$, and $-3\ln
(T+\bar T)$, so the dilaton  and the overall volume modulus couple
to all fields in the potential in a multiplicative form. If their
potential is
steep, as we have just seen, their kinetic energies will dominate
the total energy density, blocking the possibility of slow-roll
inflation. This argument was made in \cite{BS} in the context of
weakly coupled heterotic strings with field theoretic
non-perturbative potentials. Here we have extended it to all
outer regions of moduli space.

\subsection{moduli stabilization}

We would like to review here the arguments of \cite{bda1} about
moduli stabilization in the outer region of moduli space, and connect
them to our discussion about inflation. We argue that it is likely
that cosmological solutions have to end their cosmological
evolution  in the central region of moduli space. Considering only
the leading dependence of SNP effects, it is clear that the moduli are
unstable, since they have runaway potentials which force them
towards free 10D (11D) theories.
We therefore conclude
that moduli stabilization cannot occur when either the inverse
coupling or the volume are parametrically large.

It was argued in \cite{bda1} that it is very unlikely that a
minimum of vanishing cosmological constant which breaks SUSY is
found in the outer region of moduli space. But even if we assume
that such a minimum exists then there is always a deep minimum
with a  large negative cosmological constant towards weaker
coupling and larger volume \cite{BS}. In addition, there's always
a supersymmetric minimum at vanishing coupling and infinite
volume. This multi-minima structure brings into focus the
barriers separating them.  If these barriers are high enough one
may argue that flat space is a metastable state with a large
enough life time. Generically, however, this is not the case, and
classical or quantum transitions between minima are quite fast.
In the context of gaugino-condensation race-track models this was
discussed in \cite{BS}.  In particular, in a cosmological setup it
was shown \cite{BS} that classical roll-over of moduli towards
weak coupling and large volume are generic, and occur for a large
class of moduli initial conditions. Later it was shown that
cosmic friction can somewhat improve the situation
\cite{Barreiro:1998aj}, and more recently it was argued that finite
temperature effects drastically improve the situation \cite{hsow}.
The point we're making here is that the same arguments regarding
possible minima are valid
for all string theories and 11D SUGRA.

For a moment, let's consider the behavior of the complete
${\mathcal N}=1$ SUGRA action. We consider a generic moduli
chiral superfield , which we denote by $S$, which could be either
the dilaton $S$-modulus, or the $T$-modulus. We assume that its
Kahler potential is given by $K=-\ln (S+S^*)$, and that $Re S>0$,
corresponding to having a well defined compactification volume
and gauge coupling. The generic feature of the superpotential
$W(S)$ near the boundaries of moduli space is its steepness, as
described previously.  This requires that derivatives of the
superpotential are large. In mathematical terms, the steepness
property of the superpotential is expressed as follows,
 \begin{equation}
 \frac{\left|(S+S^*)\partial_S^{(n+1)} W\right|}
 {\left| \partial_S^n W \right|}\gg 1\ \ \
 n=0,1,2,3 \quad .
  \label{a20}
  \end{equation}
This property  is generic to all models of
stabilization around the  boundaries of moduli space. The typical
example of a superpotential satisfying (\ref{a20}) is a sum of
exponentials $W(S)=\sum_i e^{-\beta_i S}$, with $Re \beta_i \gg
1$, in the region $|S\beta_i| \gg 1$, the brane-instanton potentials
considered previously are clearly of this form. In this example the
``boundary region of moduli space" is simply the region $Re S \gaq
1$, but in general, the precise definition will depend on the
details of the model. It is good to keep this example in mind
while going through the following arguments, but we will not use
any particular specific form for $W$.

If one tries to stabilize moduli by introducing a more
complicated hidden sector as in race-track models
\cite{kras,deCarlos:1992da,Dine:1999dx},
then SUSY is broken at some intermediate scale (about
$10^{13}GeV)$. Alternatively, if one uses  string theoretic
mechanisms, such as the Scherk-Schwarz mechanism with the radius
of compactification stabilized by quantum effects, or models with
D branes and anti D branes at some orbifold fixed points or
D-branes at angles, SUSY is broken at the string scale. In such
mechanisms one invariably encounters what we have called the
``practical cosmological constant problem"  \cite{bda1}.
That is the problem of ensuring that to a given accuracy within a
given model the cosmological constant vanishes. This is
equivalent to the requirement that models should at the very
least allow for the possibility of a large universe to exist with
reasonable probability. This is not the same as requiring a
solution to the ``cosmological constant problem"
\cite{weinberg,carroll}:
why is the cosmological constant so small in natural units?

Thus, any cosmological solution that admits a reliable analysis in
the outer region has two striking and related problems.  One is
that the currently available stringy SUSY breaking and moduli
stabilization mechanisms are simply not viable.  The other is
that one is unable to produce sufficient slow-roll inflation to get
agreement with observation. One could take the point of view that
the difficulties are ``technical", and that they will eventually
be resolved when computational technology improves, and therefore
simply ignore them. However, we believe that the difficulties are
not technical.  Rather we believe that this indicates that
interesting cosmology and phenomenology {\em must} take place in
the central region.  Although direct results are hard to come by
in this region, we find \cite{center} that a very simple and
consistent picture does emerge, where a moduli stabilizing
potential is also a natural source for slow-roll inflation.  We
shall review this argument in section VI.

\section{Outer region Brane world cosmologies, and other models.}

Recently, several cosmological models which could perhaps be
realized in string theory have been proposed.  Among them are:
\begin{itemize}
\item
Models associated with a brane-world picture of the universe
(see, for example, \cite{Quevedo:2002xw}). We
discuss these at length below .
\item
Models with tachyonic matter and tachyon condensation
\cite{Sen:2002nu}.
This doesn't technically fit with our discussion of moduli
scalars. However, Kofman and Linde \cite{Kofman:2002rh} have shown that such
models are problematic in producing realistic inflation.
\item
Models with low string scale
(see, for example, \cite{Quevedo:2002xw}). These assume, implicitly or
explicitly, that certain moduli are stabilized in the outer region
of moduli space. Within our framework they could not be realized.
Without a new mechanism for moduli stabilization, these models
must be non-supersymmetric, and have to face the ``practical cosmological
constant problem".

\item
Time dependent orbifolds and other exact cosmological string
backgrounds (see, for example, \cite{Liu:2002ft,Elitzur:2002rt}).
These are useful in addressing conceptual
issues of time-dependent string theory, but when used to get
predictions for cosmology, should run into the usual issues that
we have presented and reviewed.

\end{itemize}

Let us discuss in more detail the brane world models.
There are two main categories
of models in the literature.

\begin{enumerate}
\item
Models in which the moduli are assumed to be fixed by some
unknown mechanism and the motion of branes (due to some mechanism
of SUSY breaking) causes inflation.
\item
Models in which the branes are at fixed points of some
orbifold/orientifold symmetry and the attraction between them
generates a potential for the moduli.
\end{enumerate}
The works
\cite{Dvali:1998pa,Burgess:2001fx,Ekp1,Dvali:2001fw,Herdeiro:2001zb}
belong to the first category while
\cite{Burgess:2001vr,Ekp2,Steinhardt:2001st}
belong to the second category. The paper of Blumenhagen et al
\cite{Blumenhagen:2002ua} has a discussion of both types of
models and in fact highlights many of the problems of these
scenarios that we discuss below. We would like to emphasize that
the assumption that the dilaton is fixed by some unknown
mechanism seems to be common to all of these models. As we have
argued previously, for consistency, this necessarily forces these
models to be central region models, since we have argued that the
dilaton cannot be stabilized in the outer region. However, this
seems to be ignored in most of the literature.

Each category can be subdivided further.

1. Fixed moduli:

\begin{itemize}
\item
\( D\bar{D} \) models: One set of models uses a brane-world which
has D branes and anti-D branes. The ten dimensions of string
theory are compactified on a torus. The branes are slowly
attracted to each other, generating inflation, provided that one of
them is placed at a Lagrange point in the compact space (the
center of the lattice generating the torus) and the other at a
corner. The center is an equilibrium point because of the mirror
images. Inflation ends when the distance between the branes
becomes less than the string scale and the open string joining
the brane becomes tachyonic, resulting in the annihilation of the
\( D\bar{D} \) system. In addition to the generic problems
discussed below this scenario has the problem of finding a set up
where some branes remain so that it could be identified with
our world (see, for example, \cite{DvalVil}). Also, of course, the
initial conditions are fine tuned.
\item
More complicated brane constructions: Here supersymmetry is
broken by considering $D_pD_{p+2}$ systems or D branes with
magnetic fields with other branes (NS and D type) suspended
between them. Here although there is no problem with identifying
the world we live in, various combinations of scalar fields need
to be fixed by hand. A detailed critique of these models is
contained in \cite{Blumenhagen:2002ua}.

\item
Ekpyrotic version I \cite{Ekp1}. In this one starts with
the Horava-Witten theory with five branes parallel to the wall.
The model is compactified on a six manifold and SUSY is broken by
a small amount. These are supposed to be alternatives to
inflation and requires that the branes be parallel to a high
degree of accuracy in spite of the SUSY breaking. For a critical
discussion of this model see \cite{Kallosh:2001du}.
\end{itemize}
All of these models have to face the moduli stabilization
problem. Additionally, potentials generated in the outer region
will not be of the slow-roll type and the question of whether or
not there is inflation is determined by the moduli. In other
words, the only way that the D brane motion can be responsible
for slow-roll inflation is if the rolling of the moduli (for
example, the size of the compact space) is slower than that of
the open string moduli, or if the moduli are stabilized at a
scale much larger than the string scale. This appears to be very
unlikely as we have argued at length earlier. In the case of the
Ekpyrotic model this would mean stabilizing the length of the
11th dimension at values larger than the M theory scale and this
too is unlikely since in this region there is a runaway potential
that tends to send this interval to infinity.
One interesting test that we leave to future investigation with a more
tightly
constrained model for the moduli superpotential is to look at a
brane-world background
that is capable of producing both brane-induced and modular
inflation, and to see in detail how these two mechanisms compete,
thus testing some of the assumptions used in brane-world models with
fixed moduli.

2. Moduli as inflatons:

\begin{itemize}
\item An alternative to the \( D\bar{D} \) annihilation picture is one
in which we have fractional branes that are forced to remain at
the fixed points of the orbifold symmetry\cite{Burgess:2001vr}.
In this construction one can achieve tadpole cancellation without
invoking equal numbers of branes and anti-branes of the same
type. However, now the attraction between the brane-anti-brane
system results in the contraction of the compact space (torus).
The radius modulus of the torus is the inflaton. Inflation ends
as before when this radius becomes smaller than the string scale.
However, standard string theory arguments would indicate that as
a radius goes below the string scale the system should be
described by the T-dual theory. Otherwise one would have to
account for an infinite number of winding modes that will become
massless. In the T-dual situation the contraction below the
string scale is actually an expansion. So instead of an end to
inflation what one would have in this scenario is a universe in
which inflation alternates with deflation! In fact
\cite{Blumenhagen:2002ua} contains a detailed investigation of
this type of scenario. As those authors point out all previous
cases are special cases of their analysis.

Assuming that the dilaton is stabilized at weak coupling (see
above) what is found is that the question of whether an
inflationary cosmological scenario is obtained or not depends on
the coordinates that are used to parametrize the Kahler and
complex structure moduli of the compact space, and in particular
on the fields that are assumed to be fixed during the
inflationary era. In the so-called gauge coordinates one can get
inflation provided all the moduli except for one are fixed by some
unknown mechanism. On the other hand in the so-called Planck
coordinates it is not possible to have any inflation. We believe
that this should be interpreted to mean that such models of outer
region brane cosmology are not very useful in trying to get
generic M-theoretic cosmological scenarios.
\item
Ekpyrotic version II \cite{Ekp2}. In this
scenario, while the CY space is stabilized by some unknown
mechanism, the walls feel an attractive potential and move toward
each other. In both versions the collision is interpreted as the
big bang and one has no need for inflation (assuming that the
colliding five brane remains parallel to the wall to a high
degree of accuracy in spite of supersymmetry breaking).
\end{itemize}

The Ekpyrotic scenario is based on the Horava-Witten theory with
ten dimensional walls separated along an 11th dimension.
We would like to characterize it in the framework that we have
developed.
The theory is compactified to 4+1 dimensions on a six dimensional
Calabi-Yau space. Let us call the distance between the walls \(
\pi \rho  \) and the distance from the {}``visible{}'' wall and
the five brane (in the first version) \( Y. \) The arguments
used in the Ekpyrotic cosmological scenario entail the shrinking
of \( Y \) (in the first version) or \( \pi \rho  \) (in the
second version) to zero. The geometrical picture that is being
used however is valid only for \( Y,\pi \rho >l_{11} \) the
eleven dimensional Planck scale. For $Y<l_{11}$ but $\pi \rho>
l_{11}$ (the situation in the first version as the  5-brane and
the wall are about to collide) we are still in the 11 dimensional
(M-theory) framework, but it is not possible to make the
assumption that a low energy supergravity description is valid
for the collision. In this sense, the model describes a cosmology
that starts in the outer region of moduli space before the
collision occurs and goes towards the central region. Lacking a
microscopic M-theory description it is not entirely clear how one
can make any conclusions about this collision. In the next
section we will argue that it is possible that an inflationary
epoch starts when the branes are in the central region and
inflation ends when they stick together. The ``pre-history" is
washed out by inflation. Let us therefore move on to the second
version which  appears to be currently favored.

In this version, the boundary branes are initially far
apart, \( \pi \rho >l_{11} \) and the appropriate description is
indeed 11D supergravity compactified in this case to 5
dimensions. As in the first version, here too it
is assumed that a potential generated by membranes   or an initial
velocity will cause the branes to move toward each other.
However, now as the branes pass through to the region \( \pi \rho
<l_{11} \) (assuming that the system does not get trapped in the
region \( \pi \rho \sim l_{11} \) - more on this later) there is
a new description, namely 10 dimensional perturbative heterotic
string theory compactified on the six dimensional CY space with
the dilaton taking the place of the distance between the two
branes by the usual relation \( e^{2\phi /3}=g_{s}^{2/3}=\rho
/l_{11} \). But here the behavior should be understood in terms
of string theory.

Thus,  assuming that the moduli associated with the compact space
remain fixed, our earlier four-dimensional discussion applies.
This means that, assuming the potential is negligible, the
relevant solutions are given by those of section~III, perhaps
with the addition of radiation that was produced in the collision.
Since the branes are getting closer together, the natural
interpretation in the string region is that the coupling constant
is getting smaller (\( \phi \rightarrow -\infty  \)). Note that
since in the pre-big bang (pre-collision) period our conventions
are to take \( t<0 \), we choose the negative sign in the solution
(\ref{eq:sol1mod}). As \( t=0 \) is approached  the metric becomes
singular and the coupling vanishes.

The authors of \cite{Ekp2} argue that this singularity
must be resolved through the higher dimensional interpretation.
However, in this regime the relevant interpretation is not through
11D supergravity but through 10D string theory. So the singularity
must be resolved by string effects, as indeed was attempted later
on by \cite{Liu:2002ft}. We will not enter into issues concerning
resolution of singularities, particle production  etc., rather we
focus on the issue of moduli stabilization.

In fact, in the Ekpyrotic case, as in the \( D\bar{D} \) case
discussed earlier, a key issue is what happens as the system goes
through the  central region of moduli space. If the system does
pass through, and the distance between the branes becomes small
enough, we find ourselves in the dual heterotic description at
weak coupling, so the system cannot be stabilized there, as we
have argued previously. We have argued in great detail that the
interesting physics is likely to occur when the branes are in the
central region, that is, when they are a few string lengths
apart. The point of view that we are advocating here is that
outer region evolution will lead to a system that is trapped in
the inner region of moduli space and that a realistic cosmology
requires an investigation of models in such a region. This is
what we have done in \cite{center} and we will revisit this
discussion in the next section.

\section{Consistent cosmology in the central region of moduli space}

The scenario that we propose for moduli stabilization and SUSY
breaking is the following. The central region of moduli
space is parameterized by chiral superfields of D=4, $\cN=1$
SUGRA. They are all stabilized at the string scale by SNP effects
which allow a continuously adjustable constant in the
superpotential. SUSY is broken at an intermediate scale by field
theoretic effects that shift the stabilized moduli only by a
small amount from their unbroken minima. The cosmological
constant can be made to vanish after SUSY breaking by the
adjustable constant. We find that models of such a scenario
provide a surprisingly rich central region landscape. Although
our motivation for proposing this scenario originates from string
universality, our analysis and conclusions are valid whenever a
string theory can be approximated by an effective $\cN=1$ SUGRA
in the central region of moduli space.

The point that we have made in \cite{center} is that the mechanism
that one might expect to exist in the central region to stabilize
the moduli can also generate sufficient inflation and lead to an
acceptable CMB spectrum. Any additional inflationary mechanism is
redundant, since all of its observable effects will be hidden by
the subsequent stage of inflation. We repeat the scaling argument
here to contrast it with the arguments about the steepness of the
potential in the outer region of moduli space. Steepness
arguments have led to problems for moduli stabilization, and also
to problems with obtaining slow-roll inflation. The key point is
that in the central region the existing scaling arguments
indicate that the potentials are indeed flat enough to remove
obstructions to both. The argument was originally introduced by
Banks  \cite{banks} in the context of Horava-Witten (HW) theory
in the outer region of moduli space.

In the effective 4D theory obtained after compactification of 10D
string theories on a compact volume $V_6$, moduli kinetic terms
are multiplied by compact volume factors $M_S^8 V_6$ ($M_S$ being
the string mass), and $M_{11}^9 V_7$ in M-theory
compactifications ($M_{11}$ being the M-theory scale and $V_7$
the compact volume \footnote{Note that we are using the term
M-theory in the restricted sense of being that theory whose low
energy limit is 11 dimensional supergravity.}). The curvature
term in the effective 4D action is multiplied by the same volume
factors. We may use the 4D Newton's constant $8\pi G_N=m_p^{-2}$
\footnote{Note that $m_p$ is the reduced Planck mass $m_p=2.4
\times 10^{18}$ GeV.} to ``calibrate" moduli kinetic terms, and
determine that they are multiplied by factors of the 4D (reduced)
Planck mass squared, $ \Gamma= \int d^4x
\left\{\frac{m_p^2}{2}\sqrt{- g} R +
\frac{m_p^2}{2}\partial\psi\partial\psi\right\}$. This argument
holds for all string/$M$-theory compactifications.

Strictly speaking, the argument we have presented can be expected
to be somewhat modified in the central region, perhaps by
different factors of order unity multiplying the curvature and
moduli kinetic terms. But the effective 4D theory must always
take the form of $\Gamma$ with some SUGRA potential. Using this
scaling, we find that the typical distances over which the moduli
can move within field space, while remaining within the central
region, should be a number of order one in units of $m_p$.

Following Banks \cite{banks} we argued that the superpotential $W\sim O(M_s^3)$.
This then leads to the  overall scale of generic moduli potentials
$\Lambda^4=M_S^6/m_p^2$ in compactifications of string theories. Similarly
 $\Lambda^4=M_{11}^6/m_p^2$ in compactifications of
$M$-theory. Of course, in the central region we expect that
$M_S\sim M_{11}$ so these two are similar in order of magnitude.
With the phenomenological input of the gauge coupling at the
unification scale it was estimated in \cite{center} that $M_s\sim
M_{11}\sim 10^{17}GeV$ and $\Lambda \sim 10^{16} GeV$.

Let us consider the expected form of the potential in the central
region based on the picture discussed in \cite{bda1}. By
assumption, the moduli potential has a SUSY preserving minimum
where it vanishes in the central region. Additionally, it has to
vanish at infinity (the extreme outer parts of moduli space),
since there the non-perturbative effects vanish exponentially,
and the 10D/11D SUSY is restored. Since the potential vanishes at
the minimum in the central region and it vanishes at infinity,
its derivative needs to vanish somewhere in between. Since the
potential is increasing in all directions away from the minimum
in the central region, that additional extremum needs to be a
maximum.

This means that in any direction there is at least one maximum
separating the central region and the outer region. As we
discussed, the distance of this maximum from the minimum is a
number of order one in units of $m_p$. We know that there should
be additional minima with negative cosmological constant lying
further away from the first maximum. Note that since we are
dealing with multidimensional moduli space the maxima in
different directions do not need to coincide, so in general they
are saddle points, but we focus on a single direction in field
space. The simplest models without additional tuning of the
parameters will not have additional stable local minima with
non-vanishing energy density. This would require tuning of four
coefficients in the superpotential (or Kahler potential).

While our arguments about the potential are motivated by the
generic requirements of particle phenomenology stated earlier, it
is still useful to review what evidence there is from the
perturbative regions of string theory for such a potential. The
theory in each corner of moduli space has non-perturbative
effects that originate from various Euclidean branes wrapping on
cycles of the compact space. However, in the outer region they
will only give runaway potentials that take the theory to the
zero coupling and decompactification limit. On the other hand
every perturbative theory has strong coupling (S-dual) and small
compact volume (T-dual) partners. From the point of view of one
perturbative theory, the potential in the dual theory is trying
to send the original theory to the strong coupling and/or zero
compact volume limit. Thus it is plausible that in the universal
effective field theory in the central region (where no
perturbative calculation is valid) there are competing terms in
the potential.  If the signs of the prefactors are the same and
(as expected) are of similar magnitude, the potential would
contain a minimum at $g\sim O(1)$ and the size of the internal
manifold would be of the order of the string scale.

To illustrate this consider the S-dual type I and Heterotic
SO(32) (I-HO) theories. The S-duality relations between them are
$\phi_{I} = -\phi_{HO}$, $g_{I}=\frac{1}{g_{H}}$,
$l_{I}^{2}=g_{HO}l^{2}_{HO}$. The string coupling is related to
the expectation value of the dilaton $g= e^{\phi }$, and
$l_{I,HO}$ refers to the string length in I/HO theories,
respectively. Note that the relation between scales is defined in
terms of the expectation value of the (stabilized) dilaton.  In
this case terms in the superpotential $W$ come from a Euclidean
$D_{5}$ brane wrapping the whole compact six space, or a
Euclidean $D_{1}$ wrapping a one cycle in the compact six-space
on the type I side, and a Euclidean $F_{5}$ brane wrapping the
whole compact six space, or a Euclidean $F_{1}$ brane wrapping a
one cycle in the compact six-space on the HO side. The leading
order expressions (up to pre-factors whose size we have estimated
to be $\sim M_S^3$) for these are given in eqs.(\ref{actd5}),
(\ref{actd1}), (\ref{actf5}), (\ref{actf1}).

In the central region we then expect both these competing effects
to be present(we hope to discuss this more concretely in a future
publication). As explained in \cite{bda2}, the minimum should be
supersymmetric, $W'=W=0$, where the latter may happen, for
instance, if there is an R-symmetry under which W has an R-charge
of 2 \cite{BBSMS}.

The moduli potential originates from  brane instantons which
depend exponentially on string coupling $g$, or on the size of the
11D interval or circle $\rho/l_{11}$, and on compactification
radii $R/l_s$, as we have described. Calling them
generically $\psi$, the 4D action takes the form $ \Gamma=\int
d^4 x \left\{\frac{1}{2} m_p^2 \partial\psi
\partial\psi - \Lambda^4 v(\psi)\right\}.
$ The overall scale of the dimensionless potential was estimated previously as
$\Lambda=M_S^{3/2}m_p^{-1/2}$, and the potential $v(\psi)$ can be
approximated in the central region by  a polynomial function.
Canonically normalizing the moduli in order  to compare to
models of inflation we define $\phi= m_p\psi$, and obtain $
\Gamma=\int d^4 x \left\{ \frac{1}{2} \partial\phi \partial\phi -
\Lambda^4 v(\phi/m_p) \right\}. $

As argued in \cite{center} one additional requirement, namely that
$v''(\phi_{max})\sim 1/25$, gives a potential that is capable of
producing slow-roll inflation. This condition requires some
tuning, but it is very mild. In addition,  it is technically
natural since this number is of the same order as the gauge
coupling at that scale. With this we obtain also the right
spectrum of perturbations. To summarize, what we demonstrated in
\cite{center} is that with very mild assumptions about the
central region, where we expect the moduli to be stabilized, a
cosmological scenario that is consistent with observations can be
obtained.

\section{Conclusions}

The main conclusion of this paper is that cosmological
scenarios in outer regions of moduli space, where the coupling
is weak and the compactification volume is large, are
simply not viable. Recall that here we define viable
cosmologies as those that include an era of slow-roll
inflation, after which the moduli are stabilized and the
Universe is in a state with an acceptably small cosmological
constant. As we have argued at length in \cite{bda2} (based on
the work of many authors) it is very unlikely that the closed
string moduli can be stabilized in the outer region. Thus, any
mechanism for a viable cosmology (as defined above) such as the
motion of open string moduli which arise during the motion of
branes, for example, will be vitiated by the motion of the
closed string moduli that sends the system to weak coupling and
decompactification.

Note that moduli in the outer region can, however, provide fast
roll inflationary scenarios in the ``pre-big bang" phase. We
examined various weak coupling cosmological scenarios and found
that they were related by various dualities and that in effect
they were equivalent to the standard pre-big-bang scenario with
its attendant problems. 

On the other hand, if the moduli do make their way into the
central region, where we expect a mechanism for generating a
stabilizing potential to be present, then, with some mild
tuning it is plausible that a viable cosmology can be obtained.
Thus we claim that any additional mechanism (brane motion,
tachyon condensation, etc.) is redundant. 

Thus, in our opinion, the most promising approach for
developing a viable cosmology (as defined above) from the
fundamental principles of string theory should lie not in
constructing different exotic outer region mechanisms, but in
addressing one of the central problems of string theory, namely
the problem of moduli stabilization. Recent work of Dijkgraaf
and Vafa \cite{DV} (based on earlier work by Vafa and
collaborators) give some hope that this problem may not be as
intractable as it seemed to be. This work makes it likely that
at least holomorphic quantities (such as the superpotential) in
the central region are calculable. Combining this top down
approach with the bottom up approach discussed in \cite{bda2}
to fix possible ambiguities it may be possible to arrive at a
potential which gives the cosmology that we have outlined in
the last section. 

With a more precise characterization of the superpotential, we
will be able to address many interesting outstanding issues.  One
such issue is understanding the number of moduli whose dynamics
must be considered in the inflationary epoch, in particular to
understand if there is an estimate for this number that is generic
for different backgrounds.  Another important feature is the
typical number of extrema of the potential.  If, in any particular
direction in moduli space, there is a small number of extrema,
then our estimate for the typical width in field space holds, and
in particular our argument for topological inflation outlined in
\cite{center} will continue to hold.  A related issue is the
importance of estimating the phase space volume of regions of
space-time that end up in the central region, versus regions that
end up in the outer region.

We would like to close by stressing that the moduli we have
focused on are completely generic in string compactifications,
and that our proposal is actually fairly conservative.  The main
difficulty is that the precise features are not accessible by
direct calculations.  Thus, at this stage, the indirect methods
we are pursuing are perhaps the only way of understanding the
physics of central region moduli stabilization, which we have
shown here to have a potentially interesting application to
cosmology.

\section{Acknowledgements}
This research is supported by grant 1999071 from the United
States-Israel Binational Science Foundation (BSF), Jerusalem,
Israel.  SdA is supported in part by the United States Department
of Energy under grant DE-FG02-91-ER-40672. EN is supported in
part by a grant from the  budgeting and planning committee of
Israeli council for higher education.

\section{Appendix: Cosmological Solutions in different dimensions}
In this appendix, we relate our four dimensional solutions to
corresponding solutions in 5D, 10D, and 11D.
We focus first on the 10D
solution, which is most relevant for determining the validity of
perturbative string theory.

For the ten-dimensional solutions, we use the string-frame action
and the 10D ansatz,
\begin{equation}
ds_{10}^2 = -dt_{10}^2 + a_{10}^2(t_{10}) dx^i dx^i +
b_{10}^2(t_{10}) dy^m dy^m \quad ,
\end{equation}
where the index $i$ takes values 1 to 3, and $m$ takes values 4
to 9. With this ansatz, we get the equations of motion
\begin{eqnarray}
H_{10a}^2 + 5 H_{10b}^2 + 6 H_{10a} H_{10b} -2 H_{10a} \dot{\phi}
- 4
H_{10b} \dot{\phi} + \frac{2}{3} \dot{\phi}^2 = 0 \\
\dot{H}_{10a} + (3 H_{10a} + 6 H_{10b} - 2 \dot{\phi}) H_{10a} = 0 \\
\dot{H}_{10b} + (3 H_{10a} + 6 H_{10b} - 2 \dot{\phi}) H_{10b} = 0 \\
\ddot{\phi} + (3 H_{10a} + 6 H_{10b} - 2 \dot{\phi}) \dot{\phi} =
0 \quad ,
\end{eqnarray}
with $H_{a10}$ and $H_{b10}$ defined analogously to the 5D case.

It is possible to
relate our 4D Einstein-frame solutions to 10D string-frame
solutions by
\begin{eqnarray}
ds_{10}^2 &= &- C_{10}^2 |t|^{-\frac{\sqrt{3}}{2} p_\sigma + \frac{1}{2}
p_\phi}
dt^2 + C_{10}^2 a^2(1) |t|^{\frac{2}{3} - \frac{\sqrt{3}}{2} p_\sigma +
\frac{1}{2}
p_\phi} dx^i dx^i \nonumber \\ && + e^{\frac{1}{2 \sqrt{3}} \sigma(1) +
\frac{1}{2}
\phi(1)} |t|^{\frac{1}{2 \sqrt{3}} p_\sigma + \frac{1}{2}
p_\phi}dy^m dy^m \\
C_{10}^2 &=& e^{-\frac{\sqrt{3}}{2} \sigma(1) + \frac{1}{2} \phi(1)}
\quad .
 \end{eqnarray}
After rescaling the time coordinate by $dt_{10} = C_{10}
|t|^{-\frac{\sqrt{3}}{4}p_\sigma + \frac{1}{4}p_\phi} dt$ and setting
$t_{10} = 0$ when $t = 0$, we get 10D
solutions that look like
\begin{eqnarray}
ds_{10}^2 &=& -dt_{10}^2 + a_{10}^2(1) |t_{10}|^{2 p_{10a}} dx^i
dx^i +
b_{10}^2(1) |t_{10}|^{2 p_{10b}} dy^m dy^m \\
p_{10a} &=& \frac{\frac{4}{3} - \sqrt{3} p_\sigma + p_\phi}{4 -
\sqrt{3} p_\sigma + p_\phi} \\
p_{10b} &=&  \frac{\frac{1}{\sqrt{3}} p_\sigma + p_\phi}{4 -
\sqrt{3} p_\sigma + p_\phi} \quad ,
\end{eqnarray}
where $a_{10}(1)$ and $b_{10}(1)$ can be expressed in terms of $a(1)$,
$\sigma(1)$, and $\phi(1)$:
\begin{eqnarray}
a_{10}(1) &=& C_{10} a(1) \left[ \frac{4 - \sqrt{3} p_\sigma +
p_\phi}{4C_{10}} \right]^{p_{10a}} \\
b_{10}(1) &=& e^{\frac{1}{4 \sqrt{3}} \sigma(1) + \frac{1}{4}
\phi(1)} \left[\frac{4 - \sqrt{3} p_\sigma + p_\phi}{4C_{10}}
\right]^{p_{10b}} \quad .
\end{eqnarray}
One can check that these are solutions of the 10D equations of
motion with
\begin{eqnarray}
H_{10a} &=& \frac{p_{10a}}{t_{10}} \quad,
H_{10b} = \frac{p_{10b}}{t_{10}} \\
\phi &=& \phi_{10}(1) + \frac{4 p_\phi}{4 - \sqrt{3} p_\sigma +
p_\phi} \ln t_{10}\\
\phi_{10}(1) &=& \phi(1) + \frac{4}{4 - \sqrt{3} p_\sigma + p_\phi}
\ln{\frac{4- \sqrt{3} p_\sigma + p_\phi}{4 C_{10}}} \quad .
\end{eqnarray}
It is the 10D form of the solution that determines whether we
trust string perturbation theory; when the string frame curvature
is small in string units, the compact directions are large
compared to the string scale, and the string coupling is small,
we can follow the evolution with confidence.

For compactifications of either type IIA or Heterotic $E8 \times
E8$, in addition to T-duality transformations, we also have
decompactification limits, when $\phi \rightarrow \infty$, that
map into 5D solutions
\begin{eqnarray}
ds_{5}^2 &=& C_5^2 |t|^{-\frac{1}{2 \sqrt{3}} p_\sigma -
\frac{1}{2}p_\phi} (- dt^2 + a^2(1)|t|^{\frac{2}{3}} dx^i dx^i) +
C_5^{-4}
|t|^{\frac{1}{\sqrt{3}} p_\sigma + p_\phi} dz^2\\
\sigma_5 &=& \frac{\sqrt{3}}{2} \sigma(1) - \frac{1}{2} \phi(1) +
(\frac{\sqrt{3}}{2} p_\sigma - \frac{1}{2} p_\phi) \ln{|t|} \\
C_5 &=& e^{-\frac{1}{4 \sqrt{3}} \sigma(1) - \frac{1}{4} \phi(1)} \quad ,
 \end{eqnarray}
which become, after rescaling the time variable by $dt_{5} = C_5
|t|^{-\frac{1}{4 \sqrt{3}} p_\sigma - \frac{1}{4} p_\phi} dt$,
\begin{eqnarray}
ds_5^2 &=& -dt_5^2 + a_5^2(1)|t_5|^{2p_{5a}} dx^i dx^i +
c_5^2(1)|t_5|^{2p_{5c}} dz^2 \\
\sigma_5 &=& \sigma_5(1) + p_{5\sigma}\ln{|t_5|} \quad .
\end{eqnarray}
We find
\begin{eqnarray}
p_{5a} &=& \frac{ \frac{4}{3} - \frac{1}{\sqrt{3}}p_\sigma -
p_\phi}{4 -
\frac{1}{\sqrt{3}} p_\sigma - p_\phi} \\
p_{5b} &=& \frac{ \frac{2}{\sqrt{3}}p_\sigma + 2 p_\phi}{4 -
\frac{1}{\sqrt{3}} p_\sigma - p_\phi} \\
p_{5\sigma} &=& \frac{ 2 \sqrt{3}p_\sigma - 2 p_\phi}{4 -
\frac{1}{\sqrt{3}} p_\sigma - p_\phi} \\
a_5(1) &=& a(1) C_5 \left[\frac{4 - \frac{1}{\sqrt{3}} p_\sigma -
p_\phi}{4
C_5} \right]^{p_{5a}} \\
c_5(1) &=& C_5^{-2} \left[\frac{4 - \frac{1}{\sqrt{3}} p_\sigma -
p_\phi}{4
C_5} \right]^{p_{5c}} \\
\sigma_5(1) &=& \frac{\sqrt{3}}{2} \sigma(1) - \frac{1}{2}
\phi(1) + p_{5\sigma}\ln{\left[\frac{4 - \frac{1}{\sqrt{3}}
p_\sigma - p_\phi}{4 C_5} \right]} \quad ,
\end{eqnarray}
which is thus a Kasner-type solution with $H_{5a} = p_{5a}/t$ and
$H_{5c} = p_{5c}/t$.  The 5D solution is of interest whenever all
relevant length scales are much larger than the 5D Planck scale,
giving us a 5D interpretation of cosmology.

In the 5D we have Einstein gravity coupled to one scalar
$\sigma_5$ as in eqs.(\ref{eq:5limit1},\ref{eq:5limit2}). With
the metric ansatz
\begin{equation}
ds_5^2 = -dt_5^2 + a_5^2(t_{5}) dx^i dx^i + c_5^2(t_{5}) dz^2
\quad ,
\end{equation}
the corresponding equations of motion are
\begin{eqnarray}
H_{5a}^2 + H_{5a} H_{5c} = \frac{1}{12} \sigma_5^2 \\
\dot{H}_{5a} + (3 H_{5a} + H_{5c}) H_{5a} = 0\\
\dot{H}_{5c} + (3 H_{5a} + H_{5c}) H_{5c} = 0\\
\ddot{\sigma}_5 + 3 H_{5a} \dot{\sigma}_5 + H_{5c} \dot{\sigma}_5
= 0 \quad ,
\end{eqnarray}
where a dot denotes the derivative with respect to $t_5$, $H_{5a}
= \dot{a}_5/a_5$, and $H_{5c} = \dot{c}_5/c_5$. It is possible to
verify directly that the mapped solutions that we have found
above are indeed solutions of the equations of motion.

Our solution can also
be interpreted by the 11D limit
 \begin{eqnarray} ds_{11}^2 &=&
C_{11}^2 |t|^{-\frac{\sqrt{3}}{2}p_\sigma - \frac{1}{6} p_\phi}
(-dt^2 + a^2(1)|t|^{\frac{2}{3}}dx^i dx^i) \nonumber \\ && +
C_{11}^2 e^{\frac{2}{\sqrt{3}}\sigma(1)} |t|^{\frac{1}{2\sqrt{3}}
p_\sigma - \frac{1}{6} p_\phi}dy^m dy^m
 + e^{\frac{4}{3} \phi(1)} |t|^{\frac{4}{3}
p_\phi} dz^2 \\
C_{11} &= &e^{- \frac{\sqrt{3}}{4} \sigma(1) - \frac{1}{12} \phi(1)} \quad
.
 \end{eqnarray}
After rescaling the time coordinate by $dt_{11} =
C_{11}|t|^{-\frac{\sqrt{3}}{4}p_\sigma - \frac{1}{12}p_\phi}dt$ this
becomes
\begin{equation}
ds_{11}^2 = -dt_{11}^2 + a_{11}^2(1)|t|^{2p_{11a}} dx^i dx^i + b_{11}^2(1)
|t|^{2p_{11b}} dy^m dy^m + c_{11}^2(1) |t|^{2p_{11c}}dz^2 \quad ,
\end{equation}
which is a Kasner type solution with $H_{11a,b,c} = p_{11a,b,c}/t_{11}$
with the various constants determined as
\begin{eqnarray}
p_{11a} &=& \frac{4-3 \sqrt{3} p_\sigma - p_\phi}{12 - 3 \sqrt{3}
p_\sigma -
p_\phi} \\
p_{11b} &=& \frac{3 \sqrt{3} p_\sigma - p_\phi}{12 - 3 \sqrt{3}
p_\sigma -
p_\phi} \\
p_{11a} &=& \frac{8 p_\phi}{12 - 3 \sqrt{3} p_\sigma - p_\phi} \\
a_{11}(1) &=& C_{11} a(1) \left[\frac{12 - 3 \sqrt{3} p_\sigma -
p_\phi}{12C_{11}} \right]^{p_{11a}} \\
b_{11}(1) &=& C_{11} e^{\frac{1}{\sqrt{3}}\sigma(1)}
\left[\frac{12 - 3 \sqrt{3} p_\sigma -
p_\phi}{12C_{11}} \right]^{p_{11b}} \\
c_{11}(1) &=& e^{\frac{2}{3} \phi(1)} \left[\frac{12 - 3 \sqrt{3}
p_\sigma - p_\phi}{12C_{11}} \right]^{p_{11c}} \quad .
\end{eqnarray}
We trust the 11D decompactification limits whenever the
curvature is lower than the 11D Planck scale and any compact
directions are larger than 11D Planck scale.

It is possible to verify
that the above solution is indeed
a solution of 11D Einstein gravity with a
metric ansatz
\begin{equation}
ds_{11}^2 = -dt_{11}^2 + a_{11}^2(t_{11}) dx^i dx^i +
b_{11}^2(t_{11}) dy^m dy^m + c_{11}^2(t_{11}) dz^2 \quad .
\end{equation}
which leads to the following equations of motion
\begin{eqnarray}
H_{11a}^2 + 5 H_{11b}^2 + 6 H_{11a} H_{11b} + H_{11a} H_{11c} + 2
H_{11b}
H_{11c} = 0 \\
\dot{H}_{11a} + (3 H_{11a} + 6 H_{11b} + H_{11c}) H_{11a} = 0 \\
\dot{H}_{11b} + (3 H_{11a} + 6 H_{11b} + H_{11c}) H_{11b} = 0 \\
\dot{H}_{11c} + (3 H_{11a} + 6 H_{11b} + H_{11c}) H_{11c} = 0 \quad .
\end{eqnarray}

Note that although the coordinates $x^i$, $y^m$, and $z$ are the
same in each of the metrics, one has to redefine the time
coordinate appropriately for the forms assumed above.

\end{document}